\documentclass[aps,twocolumn,showpacs,preprintnumbers,prl]{revtex4}
\usepackage{graphicx}% Include figure files
\usepackage{dcolumn}% Align table columns on decimal point
\usepackage{bm}% bold math        

\begin{document}
\title{Plasmon bands in metallic nanostructures}
\author{ J.E. Inglesfield$^1$, J.M. Pitarke$^{2,3}$, and R. Kemp$^1$}
\affiliation{
$^1$ Department of Physics and Astronomy, Cardiff University, 
Cardiff, CF24 3YB, UK\\
$^2$Materia
Kondentsatuaren Fisika Saila, Zientzi Fakultatea, Euskal Herriko
Unibertsitatea,\\ 644 Posta kutxatila, E-48080 Bilbo, Basque Country\\
$^3$Donostia International Physics Center (DIPC) and Centro
Mixto CSIC-UPV/EHU,\\
Manuel de Lardizabal Pasealekua,
E-20018 Donostia, Basque Country}       

%\date\today
\date{January 15, 2004}

\begin{abstract}
The photonic band structure of a three-dimensional lattice of metal
spheres is calculated using an embedding technique, in the frequency 
range of the Mie plasmons. For a small filling factor of the spheres,
Maxwell-Garnett theory gives an almost exact description of the dipole
modes, and the multipole modes are fairly dispersionless. For a larger
filling factor, crystal field effects modify the multipole
frequencies, which show dispersion. These multipole bands are enclosed
between the dipole modes. For touching spheres, there is a wide continuum
of plasmon modes between zero frequency and the bulk metal plasmon
frequency, which yield strong absorption of incident light. These plasmon 
modes are responsible for the blackness of colloidal silver.
\end{abstract}


\maketitle

An isolated free-electron metallic sphere shows Mie plasmons at 
frequencies
\begin{equation}\label{mie}
\omega_l=\omega_p \sqrt{\frac{l}{2l+1}},
\end{equation} 
where $\omega_p$ is the bulk plasmon frequency, and $l$ is the
quantum number identifying the multipole. This assumes the Drude
dielectric function,
\begin{equation}
\epsilon(\omega)=1-\frac{\omega_p^2}{\omega(\omega+i/\tau)},
\end{equation}
where $\tau$ is the scattering time. In a lattice, crystal field
effects shift the Mie plasmon frequencies, an effect which has been
studied both experimentally and theoretically~\cite{bergman}. These theoretical 
studies have generally been restricted to the dipole plasmons and 
their crystal field, but recently a number of methods have been developed for
a full solution of Maxwell's equations in periodic
structures~\cite{yab,soukoulis,pendry0,abajo,ingles}. There have been photonic band
structure calculations of lattices of metallic rods~\cite{english},
cylinders~\cite{moroz1}, and spheres~\cite{pendry,modinos,moroz2,sheng}. The plasmon
modes and their interaction with light have been studied by Yannopapas
\emph{et al.}~\cite{modinos}, in the low filling fraction regime.

In this paper, we use the new embedding method to solve Maxwell's
equations~\cite{ingles}, and consider the evolution of plasmon bands in a lattice of nanoscale metallic spheres as the sphere size increases from very small, practically the  isolated sphere limit, to touching spheres. Our band structures and 
densities of states show clearly the interaction of light with such 
systems, and the way that the plasmons are affected by the crystal field. We find that the presence of structured metal introduces a continuum of plasmon modes, which yield strong absorption of incident light and are responsible for the unique optical properties of colloidal metals.

In the photonic embedding method, the dielectric objects -- here metallic
spheres -- are replaced by an embedding ``potential'' over their
surface. This is added to the wave equation for the uniform vacuum
region between the spheres, and the equation is only solved explicitly
in this region. The embedding potential, in reality a surface tensor operator,
ensures that the surface parallel components of the $\mathbf E$ and
$\mathbf H$ fields match across the boundary between the vacuum and
the metal spheres. The electromagnetic field is expanded in terms of
any suitable basis set (we use vector plane waves) in the region
between the spheres, and because this field has no discontinuities the
expansion converges very well. In the case of non-metallic dielectric spheres,
the convergence is at least an order of magnitude  better than a direct plane
wave expansion through the whole of space  (the usual method for finding
photonic band structures)~\cite{kemp};  in the difficult case of metallic
spheres, the embedding method continues to converge well. Of course
there is a price to pay, and this is the evaluation of the embedding
tensor which replaces the metallic spheres. This involves finding the exact
solution of Maxwell's equations inside the spheres for each multipole
$l$ at frequency $\omega$, but we have derived straightforward
expressions that may readily be coded.

The multipoles of the embedding tensor are cut off at a maximum
value, $l_{\mbox{\tiny max}}$, imposing a limit to the plasmon modes
on each sphere. Without this, the density of states increases without
limit as $\omega$ approaches the planar surface plasmon frequency
$\omega_p/\sqrt{2}$ (the limit of large $l$ in Eq. (1)). It is also
necessary for convergence of the plane wave expansion.
There is a physical plasmon cut-off in $l_{\mbox{\tiny max}}$ 
due to Landau damping, though this is
larger than the values we use. Controlling $l_{\mbox{\tiny max}}$ also
enables us to study the evolution of the plasmon bands in more
detail. Unlike in scattering theory where $l_{\mbox{\tiny max}}$ is
similarly imposed, all higher values of $l$ are included in our
embedding, but with a different boundary condition, that the surface 
parallel components of $\mathbf H$ vanish on the surface of the sphere.

The system we shall discuss here is a face-centred cubic (fcc) 
lattice of metallic spheres in vacuum, the spheres being described by the Drude
dielectric function of Eq. (2). The scaling property of Maxwell's equations
makes it natural to use the dimensionless reduced frequency
$\tilde{\omega}=\omega a/2\pi c$ and reduced wave-vector $\tilde{\mathbf
k}=\mathbf{k}a/2\pi$, where $a$ is the conventional lattice constant.
Throughout this paper we use the  reduced value of $\tilde{\omega_p}=0.1$;
taking $\omega_p=15$ eV for Al, this corresponds to a lattice constant of
$a=83$ \AA. The reduced lifetime $\tilde{\tau}$ is taken to be 1000, a value
which enables us to separate spectral features. We note that quantum
size effects are likely to be important in the small spheres we study
here -- this will be treated in later work.

\begin{figure}
\includegraphics[width=0.45\textwidth,height=0.3375\textwidth]{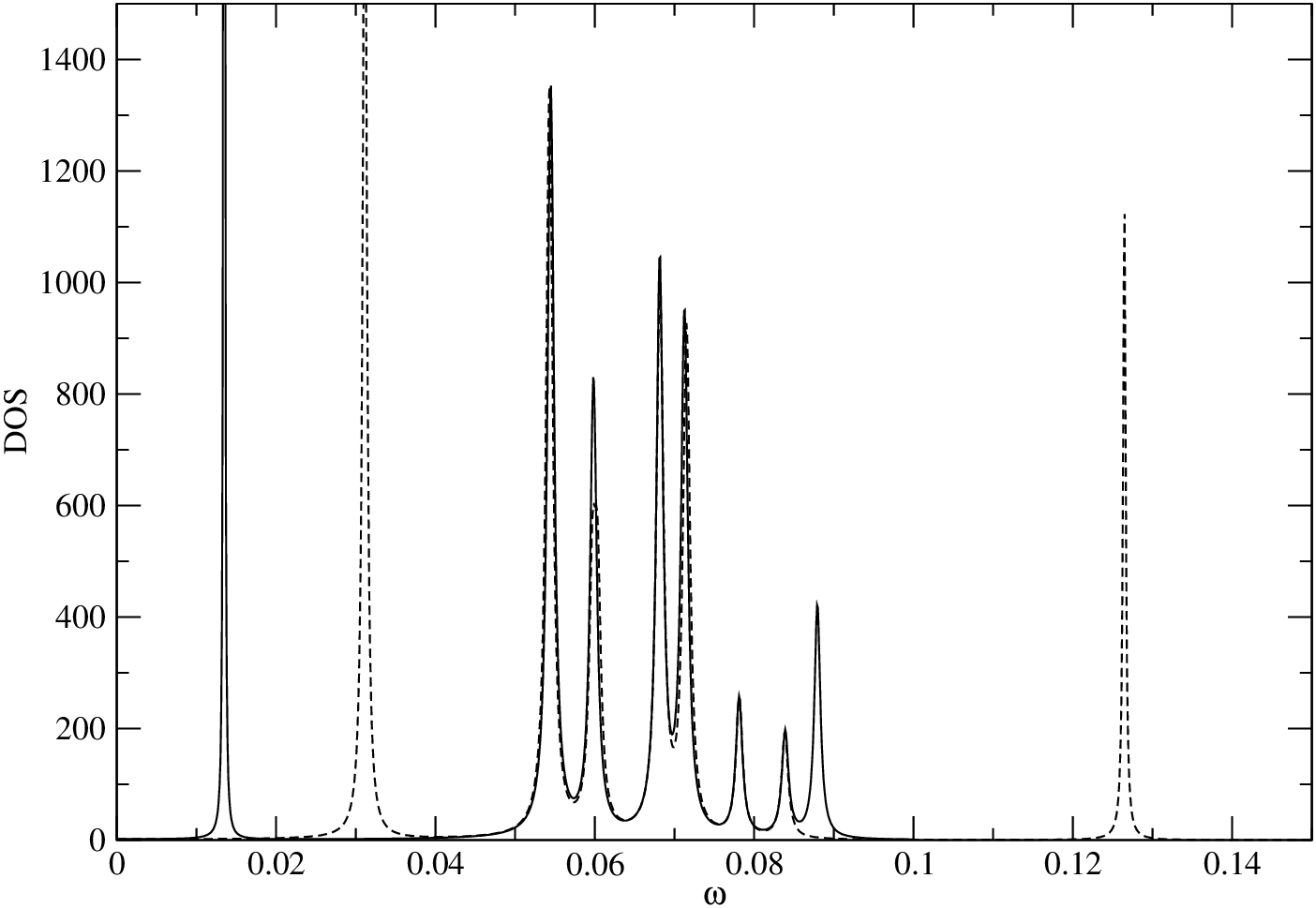}
\caption{$n_{\mathbf k}$, spectral density for fcc lattice of metal
spheres, $\tilde{r}=2$, $\tilde{\omega_p}=0.1$, $l_{\mbox{\tiny
max}}$=3, at $\tilde{\mathbf k}=(0.03,0,0)$ (solid line) and
$(0.1,0,0)$ (dashed line).\label{one}}
\end{figure}  

In the embedding method, with a frequency-dependent embedding tensor,
it is most straightforward to use a plane-wave expansion for
the Green function between the spheres, and from this we calculate the spectral
density $n_{\mathbf k}(\omega)$ (proportional to the energy density)
integrated through the vacuum region. Figure~\ref{one} shows a typical
spectral density, for an fcc lattice of  metal spheres with reduced radius
$\tilde{r}=2\pi r/a=2$, corresponding to  an actual radius of $r=26.3$\AA\ and
a filling fraction of 54\%, at $\tilde{\mathbf k}=(0.03,0,0)$  and $(0.1,0,0)$
(the X-point corresponds to $\tilde{\mathbf k}=(1,0,0)$). In this calculation 
$l_{\mbox{\tiny max}}$=3, and convergence has been achieved
with 18 transverse and 181 longitudinal
plane waves. The number of transverse waves stays very low over the
frequency range shown, but the number of longitudinal waves increases
as the radius of the sphere decreases and $l_{\mbox{\tiny max}}$
increases. The large number of longitudinal waves required is
characteristic of plasmon systems. The almost constant features of
Fig.~\ref{one} between $\tilde{\omega}=0.055$ and 0.085 are the multipole
plasmons with a  non-degenerate dipole plasmon at the top of this
frequency range. The dispersing peaks below and above this range are the light line coupled to a doubly degenerate dipole mode.

\begin{figure}
\includegraphics[width=0.45\textwidth,height=0.3375\textwidth]{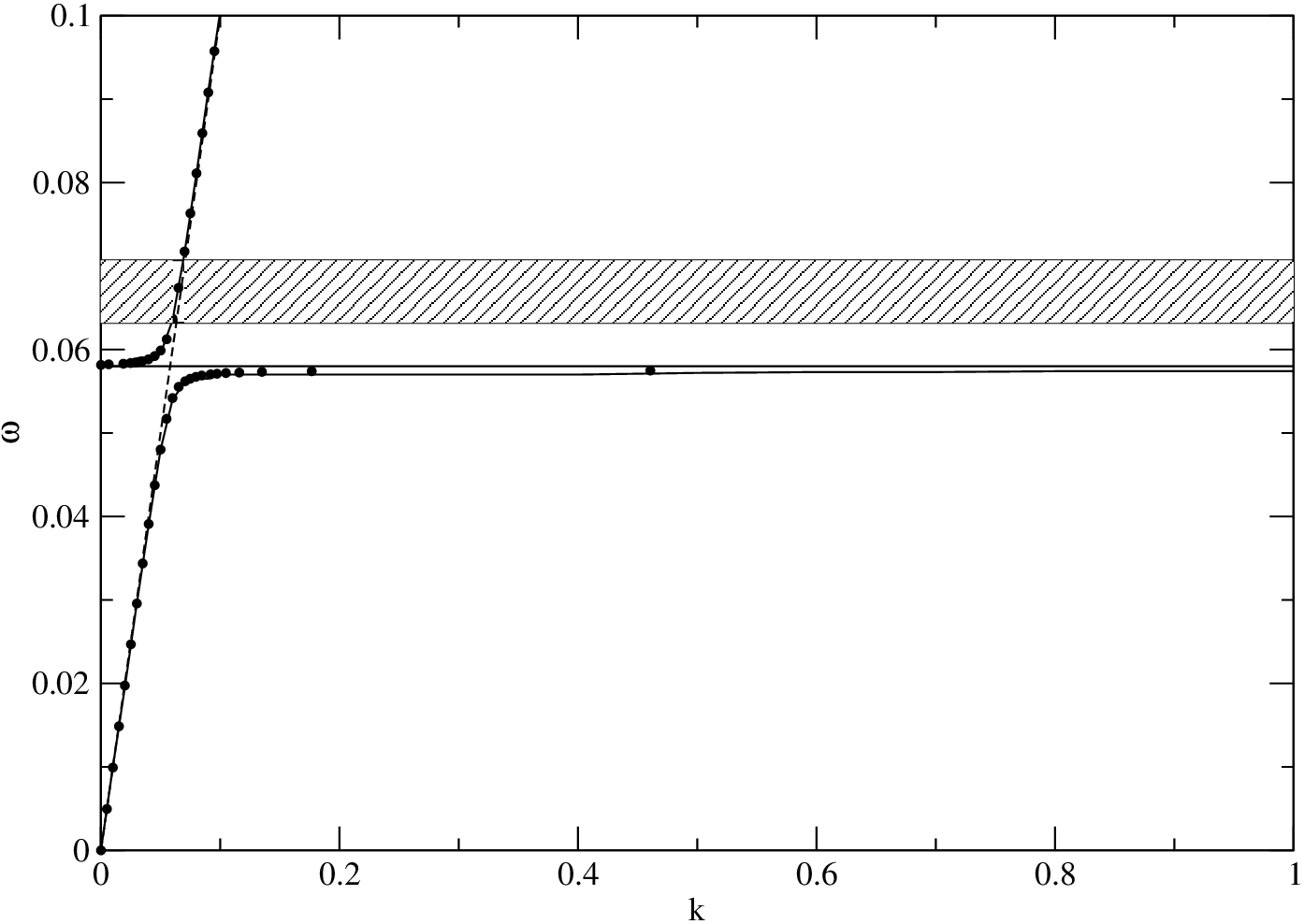}
\caption{Band structure for fcc
lattice of metal spheres, $\tilde{r}=0.5$, $\tilde{\omega_p}=0.1$, in $\Gamma$X direction (solid lines and shaded area). The dashed line represents the free-space dispersion relation $\tilde\omega=\tilde k$. Filled circles are
results from Maxwell-Garnett theory.\label{two}}
\end{figure}

From the peaks in the spectral density we plot the photonic band
structure. We begin by considering an fcc lattice of relatively
small spheres with $\tilde{r}=0.5$, an actual radius of 6.6 \AA\ and a
filling fraction of 0.8\%. The band structure for $l_{\mbox{\tiny max}}=1$ (solid lines in Fig.~\ref{two}) shows flat dipole modes at the Mie frequency of
$\omega_p/\sqrt{3}$ (solid horizontal line) and the doubly degenerate light line (solid lines below and above the flat dipole modes). At the crossing point the light lines mix with two of the dipole modes, but with small spheres the interaction and the splitting are small. This calculation required 2 transverse and 1591 longitudinal plane waves for convergence. Increasing $l_{\mbox{\tiny max}}$ does not change the light line and flat dipole modes, but gives the flat multipole bands at the frequencies given by Eq. (1) filling the shaded area of Fig.~\ref{two}.

At low filling fractions, the interaction between the spheres can
be considered within the Maxwell-Garnett approximation~\cite{mg}.
For the light mode that we see in Fig.~\ref{two}, we can define a 
frequency-dependent effective dielectric  function
\begin{equation}\label{eff}
\epsilon_{\rm eff}(\omega)={\tilde k}^2/{\tilde\omega}^2,
\end{equation}
and the Maxwell-Garnett approximation then gives~\cite{pitarke1}
\begin{equation}\label{eps1}
\epsilon_{\rm eff}(\omega)=1-f{1\over\left[1-\epsilon(\omega)\right]^{-1}-m}
\end{equation}
and
\begin{equation}\label{eps2}
\epsilon_{\rm eff}^{-1}(\omega)=1+f{1\over\left[1-\epsilon(\omega)
\right]^{-1}-n}.
\end{equation}
Here $f$ is the filling fraction, $m=(1-f)/3$, and $n=(1+2f)/3$. The filled circles in
Fig.~\ref{two} show the results for the  dispersion calculated from Eqs.~(\ref{eff}) and (\ref{eps1}). There is, not surprisingly, precise
agreement with our calculated results for the light dipole mode.

The concept of effective dielectric function is particularly useful for
understanding optical absorption, which is dictated by the poles of $\epsilon_{\rm eff}$. Conversely, the energy loss
of fast charged particles is dictated by the poles of $\epsilon_{\rm eff}^{-1}$. An inspection of Eqs.~(\ref{eps1}) and (\ref{eps2}) shows that in the isolated sphere limit ($f\to 0$) both optical absorption and energy loss occur at the Mie dipole frequency of $\omega_p/\sqrt{3}$. At finite filling fraction, the Maxwell-Garnett approximation predicts optical absorption at $\sqrt{m}\omega_p$ and energy loss at $\sqrt{n}\omega_p$. These are, respectively, the limit of the lower branch of
the light line and the starting frequency of the upper branch.

\begin{figure}
\includegraphics[width=0.45\textwidth,height=0.3375\textwidth]{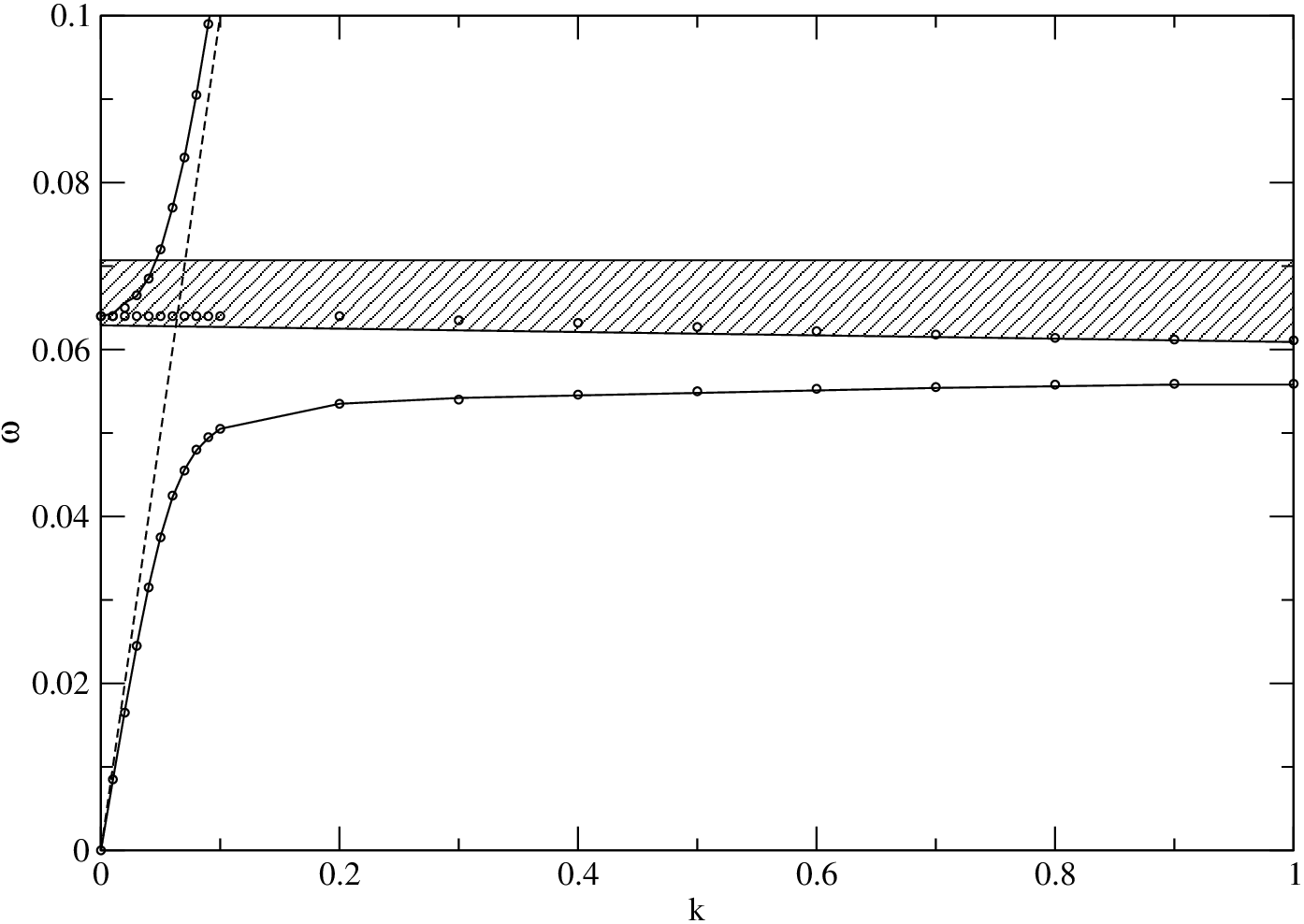}
\caption{Band structure for fcc lattice of metal spheres,
$\tilde{r}=1.2$, $\tilde{\omega_p}=0.1$, $l_{\mbox{\tiny max}}\to\infty$,
in $\Gamma$X direction (solid lines and shaded area). Circles
are results obtained with $l_{\mbox{\tiny max}}=1$. 
Maxwell-Garnett calculations, not represented here, nearly coincide
with $l_{\mbox{\tiny max}}=1$ results.\label{three}}
\end{figure}

At a sphere radius of $\tilde{r}=1.2$, filling fraction 12\%,
we find that the plasmon frequencies deviate from the isolated sphere
results. The circles of Fig. 3 show the band structure with
$l_{\mbox{\tiny max}}=1$. We see that the interaction with the light line
produces a greater splitting of the dipole modes than in Fig.~\ref{two}, which is indeed accurately reproduced by  Maxwell-Garnett theory. Increasing
$l_{\mbox{\tiny max}}$ (solid lines of Fig.~\ref{three}) barely affects the lower and upper branch of the light line, only introducing virtually flat multipole modes at the Mie frequencies of Eq.~(\ref{mie}) which fill the shaded area of Fig.~\ref{three}.

A calculation of the band structure of a simple cubic lattice of Drude
spheres with $\tilde\omega_p=0.06$ and filling fraction of $12\%$ was 
carried out by Pendry~\cite{pendry}, by using a transfer-matrix
scheme. We have calculated the band structure for those parameters, 
and have found that it has almost exactly the same form as 
the band structure of 
Fig.~\ref{three}. This disagrees, however, with the band structure 
reported in Ref.~\cite{pendry}, which shows flat low-energy bands
absent in the band structure of Fig.~\ref{three}. These probably
originate from spurious modes associated with the edges and corners
that are present in the transfer-matrix discretization procedure.

\begin{figure}
\includegraphics[width=0.45\textwidth,height=0.3375\textwidth]{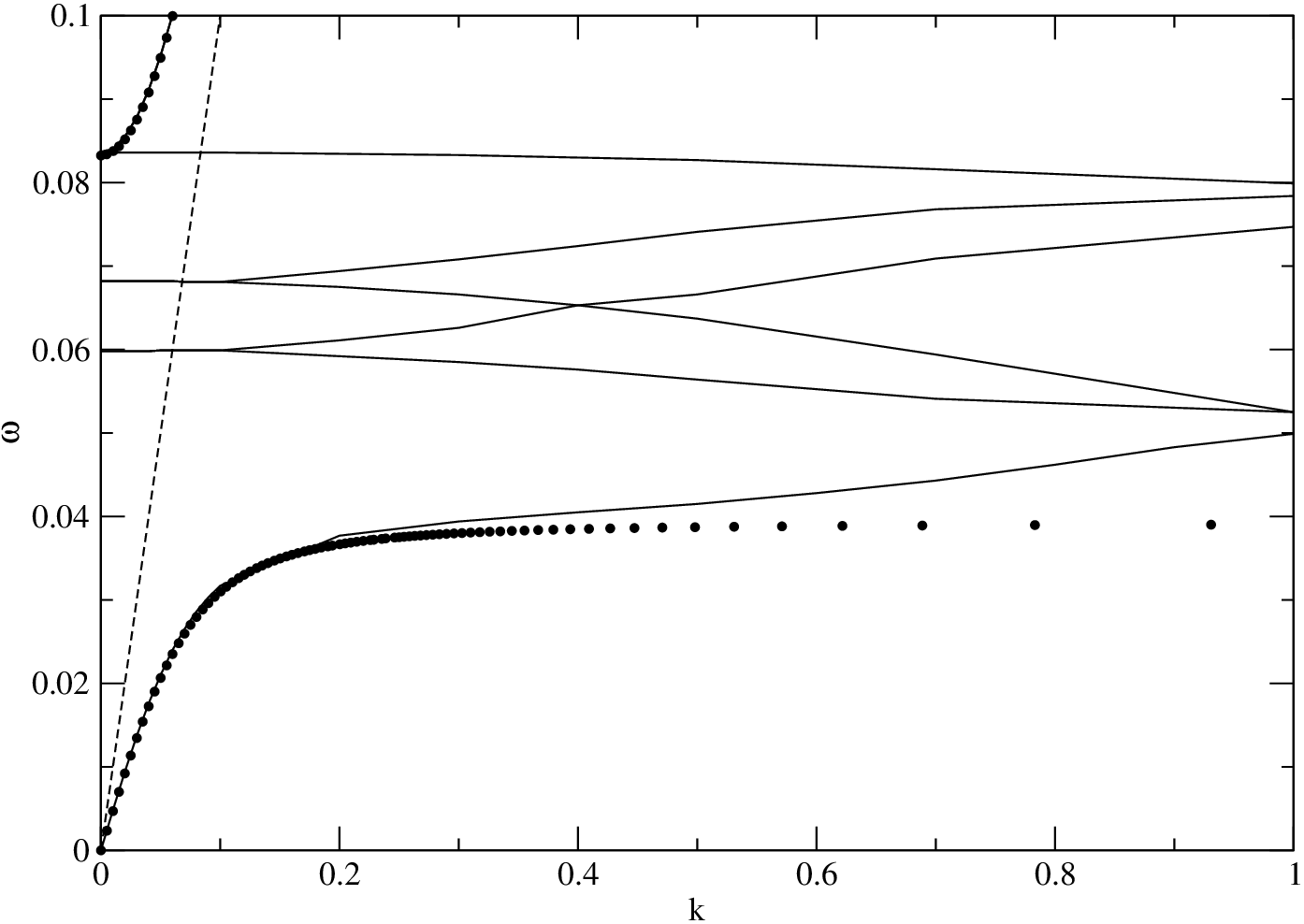}
\caption{Band structure for fcc lattice of metal spheres,
$\tilde{r}=2.0$, $\tilde{\omega_p}=0.1$, $l_{\mbox{\tiny max}}=2$,
in $\Gamma$X direction (solid lines). Filled circles are results from
Maxwell-Garnett theory.\\\\
\label{four}}
\end{figure}

\begin{figure}
\includegraphics[width=0.45\textwidth,height=0.3375\textwidth]{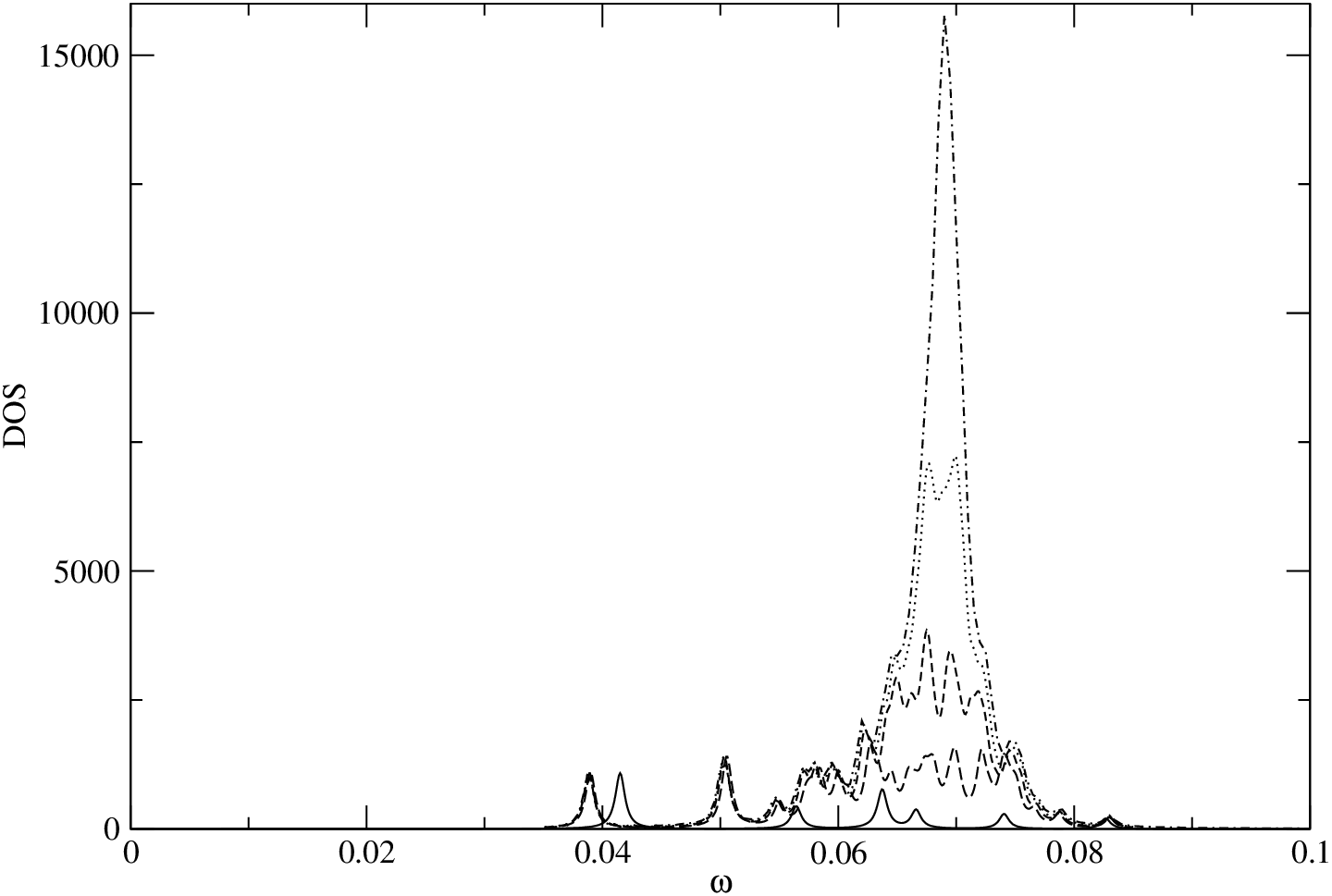}
\caption{$n_{\mathbf k}$, spectral density for fcc lattice of metal
spheres, $\tilde{r}=2$, $\tilde{\omega_p}=0.1$, at
$\tilde{\mathbf k}=(0.5,0,0)$. Solid line, $l_{\mbox{\tiny max}}=2$;
long-dashed line, $l_{\mbox{\tiny max}}=6$; short-dashed line,
$l_{\mbox{\tiny max}}=8$; dotted line, $l_{\mbox{\tiny max}}=10$; dashed-dotted line, $l_{\mbox{\tiny max}}=12$.\label{six}}
\end{figure}

\begin{figure}
\includegraphics[width=0.45\textwidth,height=0.3375\textwidth]{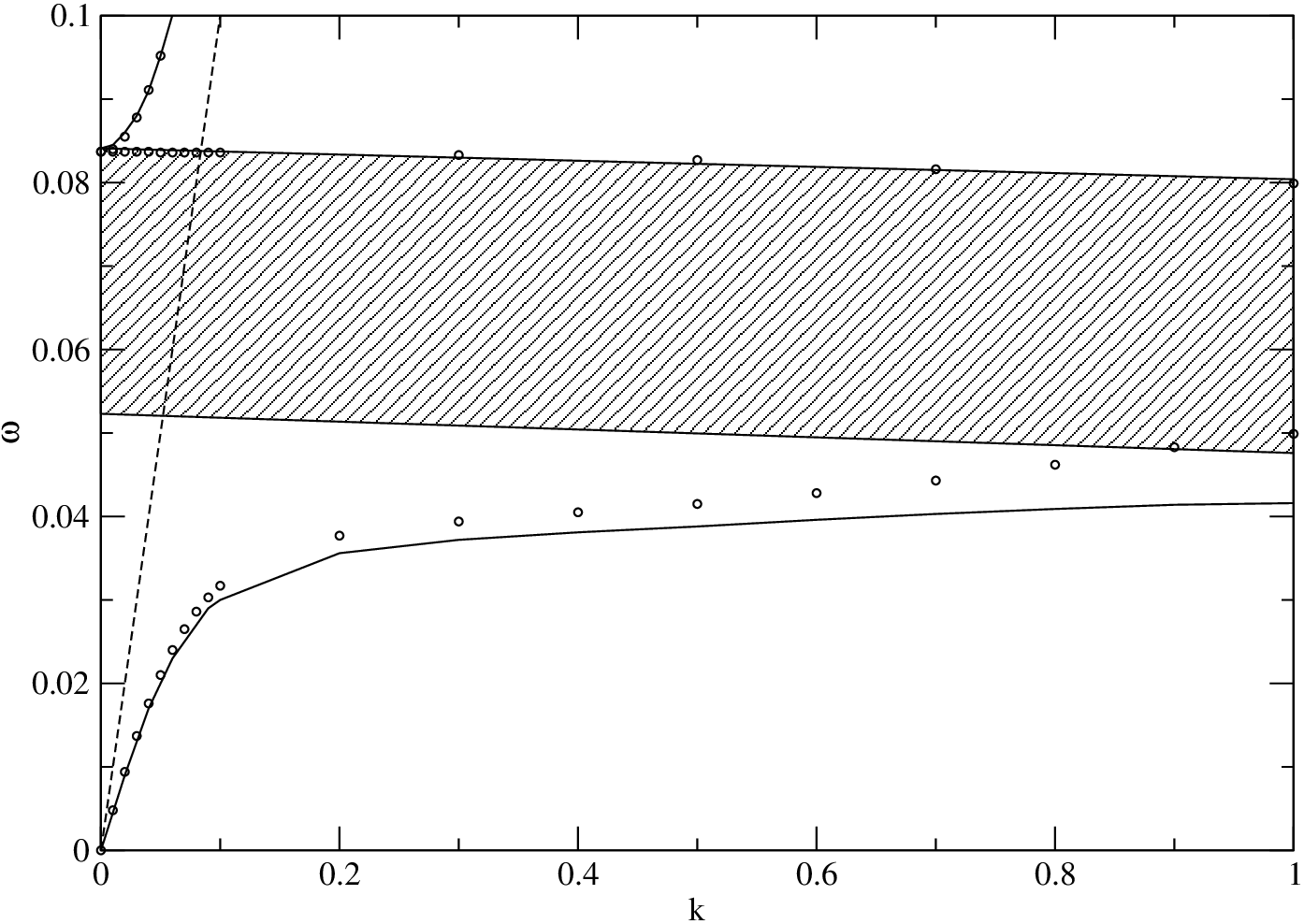}
\caption{Band structure for fcc lattice of metal spheres,
$\tilde{r}=2.0$, $\tilde{\omega_p}=0.1$, $l_{\mbox{\tiny max}}\to\infty$,
in $\Gamma$X direction (solid lines and shaded area). The circles show the dipole bands of Fig.~\ref{four}, obtained with $l_{\mbox{\tiny max}}=2$.\\\\
\label{seven}}
\end{figure}

\begin{figure}
\includegraphics[width=0.45\textwidth,height=0.3375\textwidth]{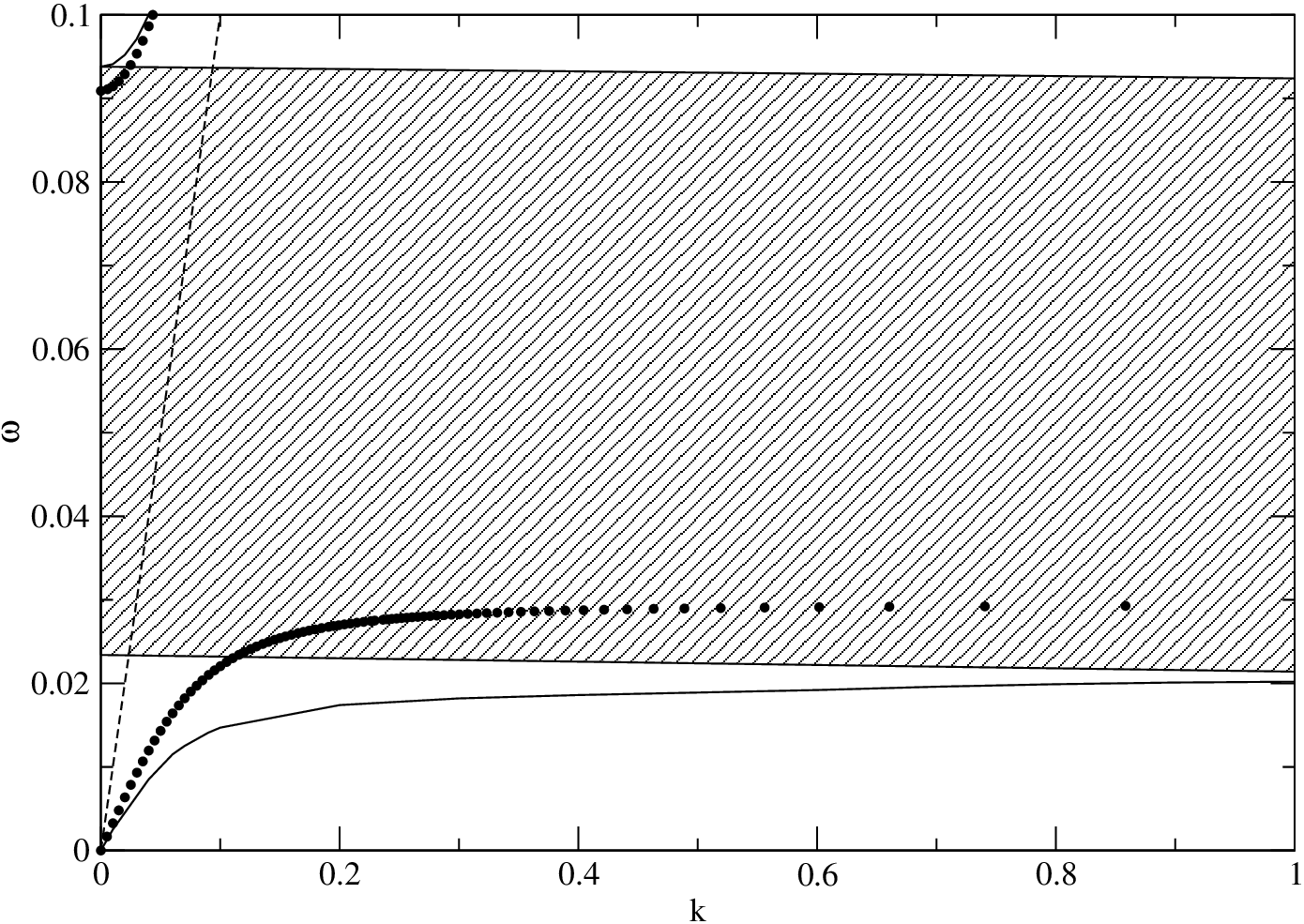}
\caption{Band structure for fcc lattice of touching metal spheres,
$\tilde{r}=2.2$, $\tilde{\omega_p}=0.1$, $l_{\mbox{\tiny max}}=12$,
in $\Gamma$X direction. Filled circles are results from Maxwell-Garnett theory.\label{eight}}
\end{figure}

The plasmon band structure is much more interesting at $\tilde{r}=2$,
filling fraction 54\%, with greater interaction between all the
modes. Figure~\ref{four} exhibits the band structure for $l_{\mbox{\tiny max}}=2$, in the $\Gamma X$ direction. We have also looked at this band structure in the
$\Gamma W$ direction, and have found that along this direction all the states are non-degenerate, with 8 plasmon bands (there is no $l=0$ plasmon). Figure~\ref{four} shows that the Maxwell-Garnett dispersion calculated from Eqs.~(\ref{eff}) and (\ref{eps1}) starts to become somewhat inaccurate, as its assumptions begin to break down. We also note that unlike in the case of smaller filling fractions, where the high multipole plasmons cut across the upper branch of the light line and lie higher than the flat dipole plasmon modes, here the dipole plasmons enclose the multipole modes. 

As $l_{\mbox{\tiny max}}$ increases, there are so many plasmon
bands that they become difficult to track individually. However, we can see the general behaviour by studying the spectral density.
Figure~\ref{six} shows the spectral density at $\tilde{\mathbf k}=(0.5,0,0)$
for $\tilde r=2$ and various values of $l_{\mbox{\tiny max}}$: 2, 6,
8, 10, and 12. We see that there is very little difference in the
overall structure for $l_{\mbox{\tiny max}}\ge 6$, the only difference being
the large increase of the spectral density associated with
high-$l$ multipole plasmons close to the planar surface plasmon limit at
$\tilde{\omega_p}/\sqrt{2}=0.07$, in the middle of the multipole
plasmon region. We can then plot the band
structure in the large $l_{\mbox{\tiny max}}$ limit, Fig.~\ref{seven} (convergence has been achieved with $l_{\mbox{\tiny max}}=8$), where the multipole bands lie in the shaded area enclosed by the dipole bands. The dipole bands
themselves have been pushed out compared with $l_{\mbox{\tiny max}}=2$ (shown in
Fig.~\ref{seven} by circles). 

It is clear from our results for $\tilde{r}=2$, showing an increase in 
dipole band gap as $l_{\mbox{\tiny max}}$ increases, that part of the 
band gap results from the interaction of the dipole modes with the
multipole modes. This interaction means that the multipole modes
can be excited by both photons and by fast charged particles. This
contrasts with the apparent absence of interaction in Fig.~\ref{three}
for $\tilde{r}=1.2$.

The behaviour of the multipoles is more extreme once the spheres
touch \cite{pack}, corresponding to $\tilde{r}=2.2$, and a filling fraction of 
74\%. Figure~\ref{eight} shows the band structure with 
$l_{\mbox{\tiny max}}=12$. The multipole modes in the shaded area are
still bounded by the dipole modes, but as $l_{\mbox{\tiny max}}$
increases the multipole modes steadily broaden, pushing the lower dipole
branch towards  $\tilde{\omega}=0$ and the upper branch towards 
$\tilde{\omega_p}$. The upper branch appears to have reached its upper
limit by $l_{\mbox{\tiny max}}=12$, just below $\tilde{\omega_p}$; the
lower branch is still moving down, and it is not clear whether this 
reaches the zero frequency of the bulk metal. However, there is
a very wide band of multipole modes nearly spanning these
limits, with corresponding optical absorption. This is the reason 
for the striking blackness of colloidal silver.

J.M.P. acknowledges partial support by the UPV/EHU, the Basque 
Unibertsitate eta Ikerketa Saila, the Spanish MCyT, and the UK EPSRC
via a visiting fellowship.

\end{document}